\documentclass[fleqn,usenatbib]{mnras}

\usepackage{newtxtext,newtxmath}

\usepackage[T1]{fontenc}
\usepackage{ae,aecompl}

\usepackage{graphicx}	
\usepackage{amsmath}	
\usepackage{amssymb}	

\title[Cross-spectrum errors]{Error formulae for the energy-dependent cross-spectrum}

\author[Adam Ingram]{
Adam Ingram$^{1}$\thanks{E-mail: adam.ingram@physics.ox.ac.uk}
\\
$ ^{1} $Department of Physics, Astrophysics, University of Oxford, Denys Wilkinson Building, Keble Road, Oxford OX1 3RH, UK
}

\date{Accepted 2019 August 28. Received 2019 August 20; in original form 2019 June 28.}
\pubyear{2019}

\begin{document}
\label{firstpage}
\pagerange{\pageref{firstpage}--\pageref{lastpage}}
\maketitle

\begin{abstract}
I present analytic error formulae for the energy-dependent cross-spectrum and rms spectrum, which are Fourier statistics widely used to probe the rapid X-ray variability observed from accreting compact objects. The new formulae cover the modulus, phase, real and imaginary parts of the cross-spectrum, and are valid for any value of intrinsic coherence between variability in different energy bands. I show that existing error formulae (including that for the phase lag), which are valid for a single cross-spectrum or power spectrum, lead to over-fitting when applied to the energy-dependent cross-spectrum - which consists of cross-spectra between individual energy channels and a common reference band. I also introduce an optimal, unbiased way to define the reference band and
{an accurate}
way to calculate the intrinsic coherence between energy bands. I find that the traditional use of the old formulae has likely had a rather benign impact on the literature, but recommend the use of the new formulae in future wherever appropriate, since they are more accurate
and are no harder to implement than existing error estimates. A code to implement the new error formulae on observational data is available online.
\end{abstract}




\section{Introduction}

Accreting compact objects exhibit rapid variability in their X-ray radiation that can be exploited in order to probe the motion of matter in strong gravitational fields and measure intrinsic properties of the compact object itself. The properties of this variability are best studied in the Fourier domain, since typically the signal to noise ratio for a given variability timescale of interest is low but an observation contains many cycles of that timescale. The most widely used Fourier statistics are the power spectrum and the cross-spectrum \citep{vanderKlis1987,vanderKlis1989}. The cross-spectrum is the product of the Fourier transform of one time series with the conjugate of the Fourier transform of another, and the power spectrum is the special case whereby the two time series are the same. It follows from the convolution theorem that the cross-spectrum is the Fourier transform of the cross-correlation function and the power spectrum is the Fourier transform of the auto-correlation function. The modulus and phase of the cross-spectrum respectively indicate the amplitude of correlated variability and the phase lag between the two time series as a function of Fourier frequency.

Fourier domain studies of black hole X-ray binaries and active galactic nuclei (AGN) reveal strong X-ray variability over several decades in Fourier frequency (e.g. \citealt{VDK2006,McHardy2006}) that is highly coherent across different energy channels \citep{Vaughan1997,Nowak1999,Uttley2014}. Phase lag measurements using the cross-spectrum reveal that, at low Fourier frequencies ($\nu \lesssim 300 ~M/M_\odot$; where $M$ is the mass of the black hole) hard photons lag soft photons. This behaviour is generally attributed to inward propagation of mass accretion rate fluctuations \citep[e.g.][]{Lyubarskii1997,Kotov2001,Ingram2013}. At higher frequencies, the phase lags between energy bands are thought to instead be dominated by light-crossing lags between directly observed rays and those that have been reprocessed by the accretion disc {(\textit{reverberation lags}; \citealt{Fabian2009,DeMarco2013,Uttley2014,DeMarco2016})}.

A particularly powerful tool is provided by the \textit{energy-dependent cross-spectrum}, which is a series of cross-spectra, each between the flux from a different energy channel and a common reference band. This enables the phase lag and correlated variability amplitude to be plotted as a function of photon energy for different frequency ranges. Related to this is the rms spectrum {(rms standing for `root mean squared')}, which is derived from the power spectrum of each energy channel integrated over a given frequency range. These tools have led to several breakthroughs, including discovery of an iron K feature in the lag energy spectrum of AGN \citep{Zoghbi2012} and X-ray binaries {\citep{DeMarco2017,Kara2019}} that is indicative of reverberation and the discovery of a quasi-periodic modulation of the iron line centroid energy indicative of relativistic precession \citep{Ingram2016}. In future, the same concept applied to X-ray polarimeters will enable searches for variability in X-ray polarization degree and angle on timescales far shorter than the minimum exposure time required for detection of polarization \citep{Ingram2017a}. As the field matures, the focus is now shifting from discovering qualitative features to measuring physical properties of the accretion flow and/or compact object by fitting theoretical models. For instance, \cite{Mastroserio2019} recently used the X-ray reverberation model \textsc{reltrans} \citep{Ingram2019} to measure the mass of the black hole in Cygnus X-1. It is therefore important to use accurate error estimates for the energy-dependent cross-spectrum in order to properly assess measurement uncertainties of model parameters.

Derivations for the error on a single power spectrum and a single cross-spectrum are presented in \citet[][originally published in 1971]{Bendat2010}. However, these formulae cannot generally be used for the energy-dependent cross-spectrum because they do not account for the use of a common reference band for all subject energy channels, and they do not account for intrinsic correlations between variability in different energy channels. This means that use of the \cite{Bendat2010} formulae for the energy-dependent cross-spectrum or the rms spectrum results in over-fitting. Here, I present analytic formulae for the error on the energy-dependent cross-spectrum and rms spectrum. These formulae are valid for {any value of} intrinsic coherence between energy channels and I demonstrate that they work even for a broad reference band constructed by summing the flux from all energy channels. In Section \ref{sec:formulae} I present the formulae, in Section \ref{sec:monte} I test the formulae with a Monte Carlo simulation and in Section \ref{sec:deriv} I present the derivations of the formulae. The logic of this structure arises from the derivation making use of asymptotic limits of the Monte Carlo simulation. In Section \ref{sec:discussion} I discuss the context in which the new formulae should be used, the optimum reference band and the effect on the literature of historical use of the old formulae. The conclusions are presented in Section \ref{sec:conclusions}. A code to implement the new error formulae on observational data with comprehensive usage instructions is available at \url{https://bitbucket.org/adingram/cross_spec_code/downloads/}

\section{Error formulae}
\label{sec:formulae}

In this section, I first quote the familiar error formulae for a single power spectrum and a single cross-spectrum. I then present new error formulae for the rms spectrum and the energy dependent cross-spectrum. These formulae take into account correlations between energy bands and the use of a single reference band for multiple cross-spectra. {All formulae quoted here assume that the observed light curves can be modelled as the sum of a signal and a noise component, such that the signal and noise contributions are completely uncorrelated with one another. This is almost always the case in X-ray astronomy, but is not generally valid (e.g. optical light curves that are heavily affected by atmospheric noise).} I first introduce the nomenclature I will use throughout.

\subsection{Definitions}

Imagine we measure the stochastically varying flux from an astronomical source in $N_e$ \textit{subject bands} and one \textit{reference band} for a time $T$. Typically the $N_e$ subject band light curves each represent the count rate in a different energy channel, and that is the context I adopt throughout this paper. Note however, that these bands could equally be the count rate for different ranges of modulation angle measured by a polarimeter such as the \textit{Imaging X-ray Polarimarey Explorer} {(\textit{IXPE}; \citealt{Weisskopf2016,Ingram2017a})}. We could imagine the reference band being the zeroth energy channel, \textit{or} alternatively it could be the sum of some or all of the subject bands. I will explore both scenarios here, and ultimately demonstrate that the two are equivalent after a very simple adjustment. The underlying power spectrum of the reference band, averaged over the $k^{\rm th}$ Fourier frequency range $\nu_k-\Delta\nu_k/2$ to $\nu_k+\Delta\nu_k/2$, is $P_r(k)$. The equivalent for the $n^{\rm th}$ subject band is $P_s(k,n)$, and the \textit{intrinsic coherence} between the $n^{\rm th}$ subject band and the reference band for the $k^{\rm th}$ frequency range is $\gamma(k,n)$. If these underlying properties remain constant over the time $T$ (\textit{stationarity}), the power spectra measured from individual segments of the light curves will be \textit{realizations} of these underlying power spectra plus a noise contribution due to photon counting statistics.

We can estimate the underlying properties by averaging over $N$ realizations. Here, the averaging can be over different segments and over different Fourier frequencies in the $k^{\rm th}$ frequency range (i.e. $\nu_k-\Delta\nu_k \leq \nu \leq \nu_k+\Delta\nu_k$), and so $N$ is the product of the number of segments in the observation and the number of frequencies in the range (see e.g. \citealt{vanderKlis1989,Uttley2014}). Denoting the $j^{\rm th}$ realization of the Fourier transforms of the flux in the reference band and the $n^{\rm th}$ subject band as $R_j(k)$ and $S_j(k,n)$ respectively, our estimates for the underlying power spectra are
\begin{eqnarray}
\tilde{P}_r(k) &=& \langle |R_{j}(k)|^2 \rangle \equiv \frac{1}{N} \sum_{j=1}^N |R_{j}(k)|^2 \label{eqn:Pr} \\
\tilde{P}_s(k,n) &=& \langle |S_{j}(k,n)|^2 \rangle \equiv \frac{1}{N} \sum_{j=1}^N |S_{j}(k,n)|^2 \label{eqn:Ps}.
\end{eqnarray}
The expectation values of these estimates are 
\begin{eqnarray}
{\rm E}\{ \tilde{P}_r(k) \} &=& P_r(k) + P_{\rm r,noise}(k) \label{eqn:Epr} \\
{\rm E}\{ \tilde{P}_s(k,n) \} &=& P_s(k,n) + P_{\rm s,noise}(k,n) \label{eqn:Eps},
\end{eqnarray}
where $P_{\rm r,noise}(k)$ and $P_{\rm s,noise}(k,n)$ are expectation values for the noise contribution in the $k^{\rm th}$ frequency range. Hereafter, a tilde will continue to denote an estimate of a variable.

In the same way, we can estimate the cross-spectrum between the $n^{\rm th}$ subject band and the reference band. For subject bands that are \textit{not} one of the channels summed in order to create the reference band light curve, the estimated cross-spectrum is
\begin{equation}
 \tilde{G}(k,n) = \langle S_{j}(k,n) R_{j}^*(k) \rangle.
 \label{eqn:rawG}
\end{equation}
For subject bands that \textit{are} included in the reference band, the estimated cross-spectrum is
\begin{equation}
 \tilde{G}(k,n) = \langle S_{j}(k,n) R_{j}^*(k) \rangle - \mathcal{N}(k,n),
 \label{eqn:subG}
\end{equation}
where $\mathcal{N}(k,n)$ accounts for the trivial correlation between the noise component of the $n^{\rm th}$ subject band and the noise component of the same subject band's contribution to the reference band. The expression for $\mathcal{N}(k,n)$ is derived in the following sub-section. It is common in the literature to avoid this trivial correlation by subtracting the subject band counts from the reference band, resulting in each subject band being crossed with a slightly different reference band \citep{Uttley2014}. I will demonstrate that this rather unsatisfying practice is not at all required, and that equation (\ref{eqn:subG}) provides a far more elegant way to deal with a broad reference band. Equations (\ref{eqn:rawG}) and (\ref{eqn:subG}) both have an expectation value of
\begin{equation}
\begin{split}
{\rm E}\{ \tilde{G}(k,n) \} = {\rm E} \left\{ {\rm Re}[\tilde{G}(k,n)] + i {\rm Im}[\tilde{G}(k,n)] \right\} \\
= G(k,n) \equiv {\rm Re}[G(k,n)] + i {\rm Im}[G(k,n)] = \gamma(k,n) S(k,n) R^*(k),    
\end{split}
\end{equation}
where $R(k)$ and $S(k,n)$ are the underlying Fourier transforms of the reference and subject bands, such that $|R(k)|^2=P_r(k)$ and $|S(k,n)|^2=P_s(k,n)$. {Note that here, following \cite{Bendat2010}, the underlying cross-spectrum is defined as $G(k,n) = \gamma(k,n) S(k,n) R^*(k)$. This is to ensure that the underlying cross-spectrum contains information on how well the light curves correlate with each other as well as the phase lag between them, plus it is mathematically convenient for the estimate of the cross-spectrum to tend to the underlying cross-spectrum as $N$ tends to infinity. The coherence is a measure of how well different realizations of the cross-spectrum line up on the complex plane (see e.g. \citealt{Nowak1999} for an intuitive discussion).} For $N \gtrsim 40$, all estimates are Gaussian distributed to a very good approximation \citep{Huppenkothen2018}. This is the limit considered throughout this paper. The most important nomenclature used throughout this paper is summarized in Table \ref{tab:defs}.

\begin{table}
\begin{center}
\begin{tabular}{||p{1.2cm}|p{6cm}||} 
\hline
Symbol  & Definition \\
\hline
\hline
$P_r$    & Reference band power spectrum \\
\hline
$P_s$    & Subject band power spectrum \\
\hline
$R$    & Fourier transform of the reference band flux time series \\
\hline
$S$    & Fourier transform of the subject band flux time series \\
\hline
$\mu_R$    & Reference band mean count rate \\
\hline
$\mu_s$    & Subject band mean count rate \\
\hline
$P_{\rm r,noise}$    & Reference band noise contribution \\
\hline
$P_{\rm s,noise}$    & Subject band noise contribution \\
\hline
$G$    & Cross-spectrum: $G={\rm Re}[G]+i{\rm Im}[G]=|G|{\rm e}^{i \phi}$ \\
\hline
$\gamma^2$  & Intrinsic squared coherence \\
\hline
$g^2$  & Raw squared coherence \\
\hline
$b^2$  & Bias term \\
\hline
$\mathcal{N}$  & See Section \ref{sec:refband} \\
\hline
\hline
Q      & Underlying value of quantity $Q$ \\
\hline
$\tilde{Q}$      & Estimate of quantity $Q$ \\
\hline
$Q_j$      & $j^{\rm th}$ realization of quantity $Q$ \\
\hline
$\langle Q_j \rangle$      & $\frac{1}{N}\sum_{j=1}^N Q_j$ \\
\hline
$dQ$      & $1\sigma$ uncertainty on quantity $Q$ \\
\hline
\end{tabular}
\end{center}
\caption{Summary of the meaning of the most frequently used symbols in this paper.}
\label{tab:defs}
\end{table}

\subsection{Calculating a cross-spectrum with a broad reference band}
\label{sec:refband}

Here, I derive the function $\mathcal{N}(k,n)$ that accounts for the trivial correlation that the subject band flux has to its own contribution to the reference band. The correct expression depends on whether the absolute or fractional rms normalization is applied. Defining the $j^{\rm th}$ realization of the reference band Fourier transform in the $k^{\rm th}$ frequency range as
\begin{equation}
    R_j(k) = \sum_{n=1}^{N_e} S_j(k,n),
\end{equation}
it is straight forward to show that the expectation value of $\langle S_{j}(k,n) R_{j}^*(k) \rangle$ is
\begin{equation}
    {\rm E}\{ \langle S_{j}(k,n) R_{j}^*(k) \rangle \} = \langle |S_{j}(k,n)|^2 \rangle + \sum_{m \neq n} S_j(k,n)S_j^*(k,m).
    \label{eqn:esr}
\end{equation}
The first term on the right hand side includes a noise term (equation \ref{eqn:Eps}) therefore, if $\tilde{G}(n,k)$ and $P_{\rm s,noise}(k,n)$ are both defined in absolute rms normalization (see e.g. Section 2 of \citealt{Ingram2013}), then
\begin{equation}
    \mathcal{N}(k,n) = P_{\rm s,noise}(k,n).
\end{equation}
{In the case of}
pure photon counting noise (i.e. {the noise is Poisson distributed with} no background subtraction, instrumental dead time etc), then $P_{\rm s,noise}(k,n) = 2 \mu_s(n)$ \citep{vanderKlis1989,Uttley2014}, where $\mu_s(n)$ is the mean count rate in the $n^{\rm th}$ subject band.

The cross-spectrum can be converted from absolute to fractional rms normalization by dividing both sides by $\mu_s(n) \mu_R$, where $\mu_R$ is the mean count rate in the reference band. The power can be converted from absolute to fractional rms normalization by dividing by $\mu_s^2(n)$. Therefore, the function that accounts for the trivial correlations becomes
\begin{equation}
    \mathcal{N}(k,n) = \frac{\mu_s(n)}{\mu_R} P_{\rm s,noise}(k,n).
\end{equation}
For pure photon counting noise, this becomes $\mathcal{N}(k,n)=2/\mu_R$, which happens to be equal to $P_{\rm r,noise}$ in fractional rms normalization.

\subsection{Error on a single power or cross spectrum}

The error on the reference band power spectrum is \citep[e.g.][]{vanderKlis1989,Bendat2010}
\begin{equation}
    d\tilde{P}_r(k) = \frac{\tilde{P}_r(k)}{\sqrt{N}},
    \label{eqn:dPr}
\end{equation}
and extension to one of the subject band power spectra is trivial. This error estimate is appropriate if one wishes to fit a model for the frequency dependence of a single power spectrum, in that the expected value of reduced $\chi^2$ for the correct model is unity. \cite{Bendat2010} show that the error for the real and imaginary parts of a single cross-spectrum is
\begin{eqnarray}
d{\rm Re}[\tilde{G}(k)] &=& \sqrt{\frac{\tilde{P}_r(k) \tilde{P}_s(k) + ({\rm Re}[\tilde{G}(k)])^2 - ({\rm Im}[\tilde{G}(k)])^2}{2N} } \nonumber \\
d{\rm Im}[\tilde{G}(k)] &=& \sqrt{ \frac{\tilde{P}_r(k) \tilde{P}_s(k) - ({\rm Re}[\tilde{G}(k)])^2 + ({\rm Im}[\tilde{G}(k)])^2}{2N} }.
\label{eqn:BPdG}
\end{eqnarray}
Again, these error estimates are appropriate if one wishes to fit a model for the frequency dependence of a single cross-spectrum, as in e.g. \cite{Rapisarda2017a}.

One may alternatively wish to fit a model for modulus, $|\tilde{G}(k)|$, and phase, $\tan\tilde{\phi}(k)={\rm Im}[\tilde{G}(k)]/{\rm Re}[\tilde{G}(k)]$, of the cross-spectrum. In this case, the \cite{Bendat2010} formulae are
\begin{eqnarray}
d\tilde{\phi}(k) &=& \sqrt{ \frac{1-\tilde{g}^2(k)}{2 \tilde{g}^2(k) N} } \\
d|\tilde{G}(k)| &=& \sqrt{ \frac{\tilde{P}_r(k) \tilde{P}_s(k)}{N} },
\label{eqn:BPdGdphi}
\end{eqnarray}
where $\tilde{\phi}(k)$ is expressed in radians and $\tilde{g}(k)$ is the \textit{raw coherence}, estimated as
\begin{equation}
    \tilde{g}^2(k) = \frac{|\tilde{G}(k)|^2 - \tilde{b}^2(k)}{\tilde{P}_r(k) \tilde{P}_s(k)},
    \label{eqn:g2}
\end{equation}
and $\tilde{b}^2(k)$ is the \textit{bias} term,
\begin{equation}
    \tilde{b}^2(k) = \frac{\tilde{P}_r(k) \tilde{P}_s(k) - \gamma^2 [\tilde{P}_r(k)-P_{\rm r,noise}(k)][\tilde{P}_s(k)-P_{\rm s,noise}(k)]}{N}.
    \label{eqn:b2}
\end{equation}
This bias term comes from the modulus of a quantity being positive definite, and therefore a noisy measurement of $|\tilde{G}|^2$ will be biased towards values slightly larger than {$|G|^2$}. Subtracting off the bias term corrects for this bias. We see from equation (\ref{eqn:b2}) that $\tilde{b}^2$ depends on the intrinsic coherence, which is not known \textit{a priori}. Usually, the bias term is calculated assuming $\gamma=1$ \citep{Vaughan1997,Uttley2014}, but note that a more general estimate can be made by setting up an iteration loop. I use such an iteration loop on \textit{Rossi X-ray Timing Explorer} (\textit{RXTE}) data in section \ref{sec:data}. For large $N$, $\tilde{b}^2$ becomes very small and it is practical to set $\tilde{b}^2=0$, which I do in sections \ref{sec:monte} and \ref{sec:deriv} in which $N=500$ is always used. Since the estimate of the bias term is itself noisy, including it even for large $N$ can lead to erroneous estimations of negative $\tilde{g}^2$. I therefore recommend always setting $\tilde{b}^2=0$ for $N\geq 500$.

\subsection{Error on the energy-dependent cross-spectrum}
\label{sec:crosserrors}

If we instead calculate $N_e$ cross-spectra with the intention of fitting a model for the energy dependence of the cross-spectrum {(\textit{the energy-dependent limit})}, we must appreciate that: i) we have used the \textit{same} reference band for each cross-spectrum, and ii) the subject bands are correlated with one another. Using the \cite{Bendat2010} formulae will therefore lead to over-fitting (i.e. the expectation value of reduced $\chi^2$ is less than unity), because these formulae account for uncertainties in the reference band that are the \textit{same} for each of the $N_e$ cross-spectra. These reference band uncertainties therefore contribute a \textit{systematic} error in that they contribute an uncertainty on the normalization of the model, but they do \textit{not} contribute to the dispersion of measurements for different subject bands. In the energy-dependent limit, the error on the real part, imaginary part and modulus of the cross-spectrum are all the same, given by
\begin{equation}
\begin{split}
    d{\rm Re}[\tilde{G}(n)] = d{\rm Im}[\tilde{G}(n)] = d|\tilde{G}(n)| = \\
    \sqrt{ \frac{\tilde{P}_r}{2N} \left[ \tilde{P}_s(n) - \frac{|\tilde{G}(n)|^2 - \tilde{b}^2(n)}{\tilde{P}_r-P_{\rm r,noise}} \right] },
    \end{split}
    \label{eqn:dG}
\end{equation}
where all dependencies on frequency range have been left as implicit for brevity (and will continue to be for the remainder of this section). I present the derivation of this equation in Section \ref{sec:dGderiv}. The error on the phase is
\begin{equation}
    d\tilde{\phi}(n) = \sqrt{ \frac{\tilde{P}_r}{2N} \left[ \frac{\tilde{P}_s(n)}{|\tilde{G}(n)|^2-\tilde{b}^2(n)} - \frac{1}{\tilde{P}_r-P_{\rm r,noise}} \right] },
    \label{eqn:dphi}
\end{equation}
and I present the derivation of this equation in Section \ref{sec:dphideriv}. In Section \ref{sec:monte}, I demonstrate that these formulae work using a Monte Carlo simulation.

It is common to define alternative statistics derived from the cross-spectrum such as the \textit{complex covariance}, $\tilde{C}(n) \equiv \tilde{G}(n) \sqrt{\Delta\nu/(\tilde{P}_r-P_{\rm r,noise})}$ \citep{Mastroserio2018} and the \textit{covariance spectrum}, $|\tilde{C}(n)|$ \citep{Wilkinson2009,Uttley2014}. The error on these quantities can be trivially obtained by multiplying equation (\ref{eqn:dG}) by $\sqrt{\Delta\nu/(\tilde{P}_r-P_{\rm r,noise})}$.

\subsection{Error on the rms spectrum}

Finally, if we calculate the power spectrum for each subject band with the intention of fitting an energy dependent model, we would again suffer from over-fitting if we used the \cite{Bendat2010} error estimate (equation \ref{eqn:dPr}). This is because the subject bands are correlated with one another, whereas using equation (\ref{eqn:dPr}) implicitly assumes that the subject bands are completely independent of one another. Instead, the correct error estimate is
\begin{equation}
    d\tilde{P}_s = \sqrt{ \frac{[1-\tilde{\gamma}^4(n)]\tilde{P}_{\rm sub}^2(n) + P_{\rm s,noise}^2(n) + 2 \tilde{P}_{\rm sub}(n) P_{\rm s,noise}(n)}{N} },
    \label{eqn:dPs}
\end{equation}
where $\tilde{P}_{\rm sub}(n) \equiv \tilde{P}_s(n)-P_{\rm s,noise}(n)$ and the estimate of the intrinsic coherence is
\begin{equation}
    \tilde{\gamma}^2(n) = \frac{|\tilde{G}(n)|^2-\tilde{b}^2(n)}{(\tilde{P}_r-P_{\rm r,noise})P_{\rm sub}(n)}.
\end{equation}
Note that for $\gamma=0$, the new formula simplifies dramatically to the old formula (equation \ref{eqn:dPr} with subscript r changed to subscript s). I present a derivation of equation (\ref{eqn:dPs}) in Section \ref{sec:dpsderiv} and demonstrate that is works with a simulation in Section \ref{sec:monte}.

It is common to define the \textit{rms spectrum}, $\tilde{\sigma}(n) \equiv \sqrt{\Delta\nu \tilde{P}_{\rm sub}(n)}$. Depending on the normalization used for the Fourier transform, this can either be used as a measure of the fractional or absolute rms variability amplitude (see e.g. \citealt{Ingram2013} or \citealt{Uttley2014} for a discussion on rms normalization). Using standard error propagation, it is straightforward to show that the error on the rms spectrum is
\begin{equation}
    d\tilde{\sigma}(n) = \sqrt{\frac{[1-\tilde{\gamma}^4(n)]\tilde{\sigma}^4(n) + \sigma_{\rm noise}^4(n) + 2 \tilde{\sigma}(n) \tilde{\sigma}_{\rm noise}(n)}{4 N \tilde{\sigma}^2(n)} },
    \label{eqn:drms}
\end{equation}
where $\tilde{\sigma}_{\rm s,noise}(n) \equiv \sqrt{\Delta\nu P_{\rm noise}(n)}$.

\section{Monte Carlo simulations}
\label{sec:monte}

In this section, I use Monte Carlo simulations to demonstrate that the formulae all work, in that they give the expected $\chi^2$ values when the simulation data are fit back with the correct model. I first assume that the reference band is entirely separate from the subject bands, before additionally simulating the case whereby the reference band is the sum of all the subject bands in order to demonstrate that the two cases are the same.

\begin{figure*}
	\vspace{-6mm}
	\includegraphics[width=\columnwidth,trim=0.5cm 1.0cm 1.5cm 0.0cm,clip=true]{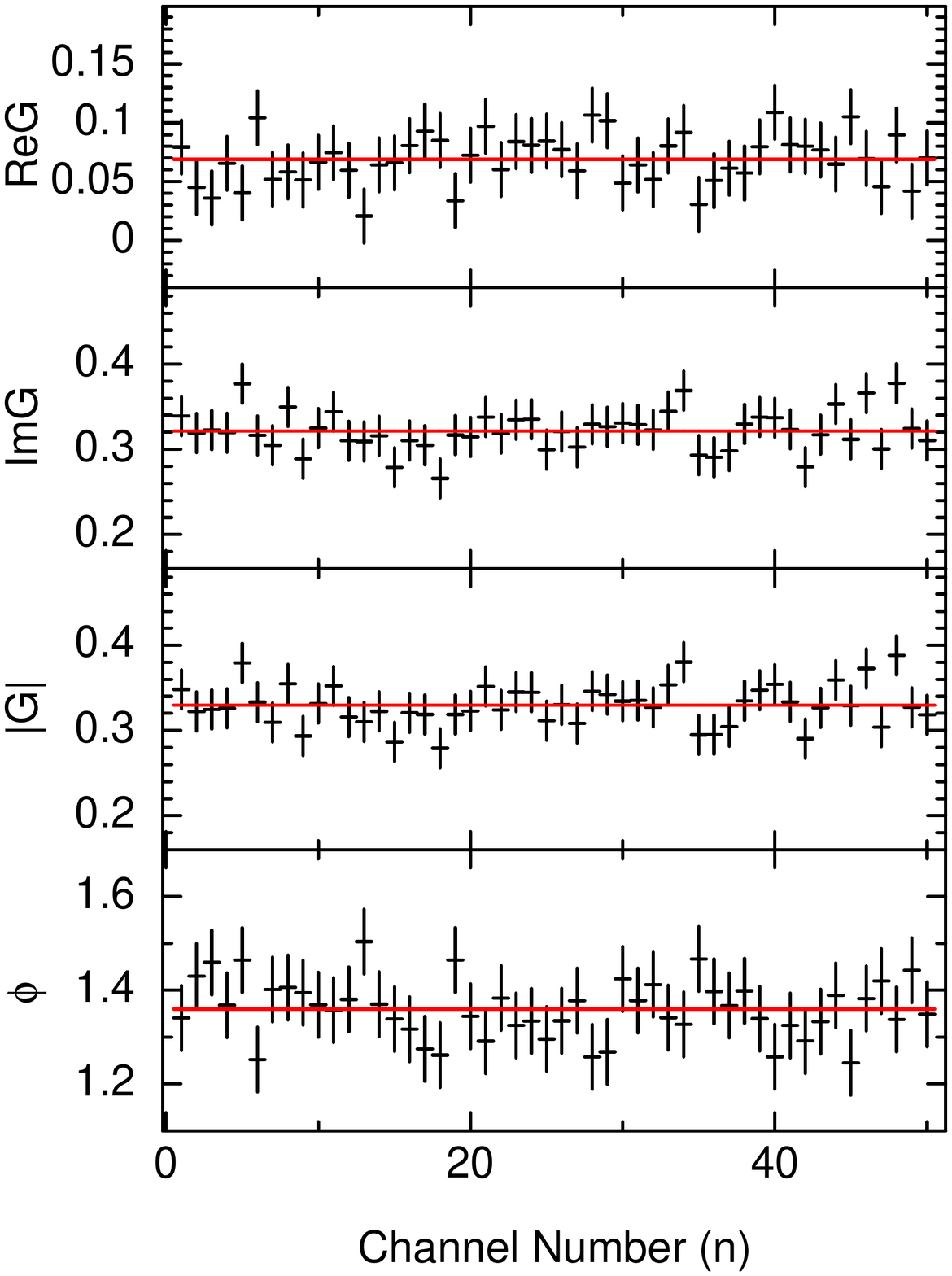} ~~~~
	\includegraphics[width=\columnwidth,trim=0.5cm 1.0cm 1.5cm 0.0cm,clip=true]{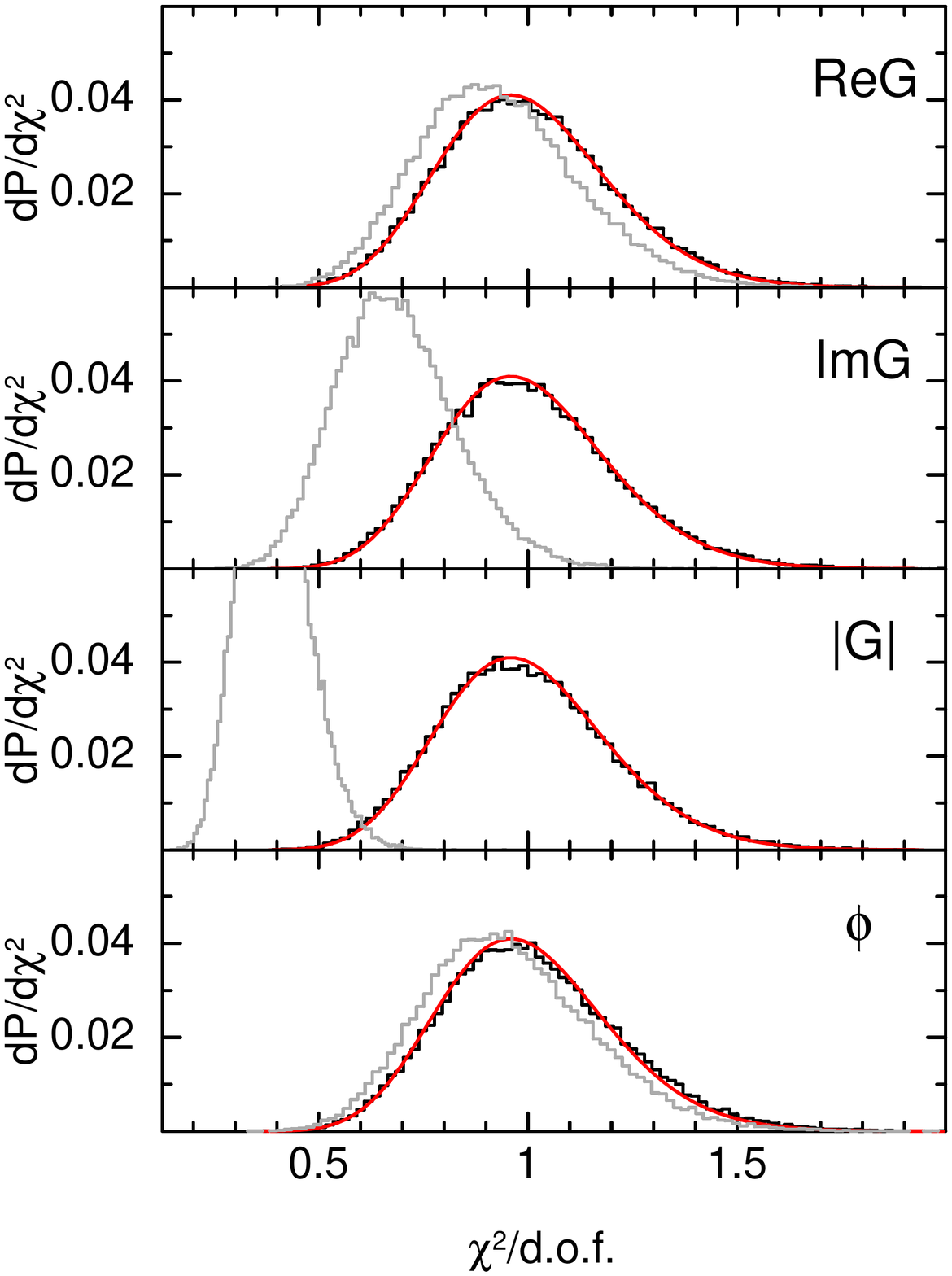} 
	\vspace{-6mm}
    \caption{\textit{Left:} Example of simulated data with error bars calculated using the analytic expressions presented in this paper (black), along with the best-fitting model (red lines). The top two panels show the real and imaginary parts of the energy-dependent cross-spectrum, whereas the bottom two panels show the modulus and phase angle (the latter is in radians). The input model does not depend on channel number and the input parameters are arbitrarily chosen: $\gamma=0.6$, $R=0.6$, $P_{\rm r,noise}=0.1$, $S(n)=0.2 + i~0.9$, $P_{\rm s,noise}(n)=0.6$, $N=500$. \textit{Right:} Probability distribution functions created by simulating 50,000 data sets (the left panel is an example of one of these data sets) and calculating the $\chi^2$ of the best-fitting model for each one (stepped lines). The black histograms are calculated using the error formulae derived in this paper, whereas the grey lines are instead calculated using the Bendat \& Piersol formulae. The red lines depict the $\chi^2$ probability distribution function with $49$ degrees of freedom ($N_{\rm e}=50$ energy channels and one free model parameter). It is clear that the new formulae reproduce the theoretical distribution very well, whereas use of the old formulae leads to over-fitting.}
    \label{fig:eg}
\end{figure*}

Let us start by defining a complex Gaussian random variable $X_{jk}$, where $j$ denotes the $j^{\rm th}$ realization and $k$ the $k^{\rm th}$ frequency range. The subscripts therefore denote that new values of the real and imaginary parts of $X_{jk}$ are drawn for each frequency range of each realization. Although the simulation only considers a single frequency range, it is instructive to carry around explicit $k$ dependencies in some of the formulae. The real and imaginary parts of $X_{jk}$ are both independent random variables drawn from a Gaussian distribution with a mean of zero and a variance of $1/2$. This means that the expectation value of $|X_{jk}|^2$ is unity. Defining another independent complex random variable $A_{jk}$ in exactly the same way, we can use these `seed' complex random variables to define the $j^{\rm th}$ realization of the Fourier transform of the reference band flux as
\begin{equation}
R_j(k) = R(k) ~ X_{jk} + \sqrt{P_{\rm r,noise}(k)}~A_{jk},
\label{eqn:Rj}
\end{equation}
where $P_r(k)=|R(k)|^2$. This is essentially the algorithm of \cite{Timmer1995}. It is fairly straight forward to show that the expectation value of $\langle |R_j(k)|^2 \rangle$ is $P_r(k)+P_{\rm r,noise}(k)$, as is required (equation \ref{eqn:Epr}). This is because $X_{jk}$ and $A_{jk}$ are uncorrelated, and so the expectation value of e.g. $\langle X_{jk} A_{jk}^* \rangle$ is zero.

To simulate the $n^{\rm th}$ subject band Fourier transform, we can define a further two seed complex random variables, with $j^{\rm th}$ realization $Y_{jkn}$ and $B_{jkn}$. These have exactly the same properties as $X_{jk}$ and $A_{jk}$, except the subscript $n$ denotes that a new value of $Y_{jkn}$ and $B_{jkn}$ is drawn for each subject band, whereas the same values of $X_{jk}$ and $A_{jk}$ are used for the $j^{\rm th}$ realization of every subject band. The $j^{\rm th}$ realization of the Fourier transform of the flux in the $n^{\rm th}$ reference band for the $k^{\rm th}$ frequency range is
\begin{equation}
\begin{split}
S_{j}(k,n) &=& S(k,n)~\left[ \gamma(k,n) X_{jk} + \sqrt{1-\gamma^2(k,n)}~Y_{jkn} \right] \\
&&+ \sqrt{P_{\rm s,noise}(k,n)}~B_{jkn},
\label{eqn:Sjn}
\end{split}
\end{equation}
where $P_s(n,k)=|S(n,k)|^2$. We can again check that this has the required properties: ${\rm E}\{ \langle |S_{j}(n,k)|^2 \rangle \} = P_s(n,k) + P_{\rm s,noise}(n,k)$ and ${\rm E}\{ \langle S_{j}(n,k)R_j^*(k) \rangle \} = G(n,k) = \gamma(n,k) |S(n,k)||R(k)|$. Because new values of $X_{jk}$ and $A_{jk}$ are not drawn for different subject bands, this simulation captures the correlations between subject bands created by using a common reference band.

\begin{figure*}
	\vspace{-6mm}
	\includegraphics[width=\columnwidth,trim=0.5cm 1.0cm 1.5cm 10.0cm,clip=true]{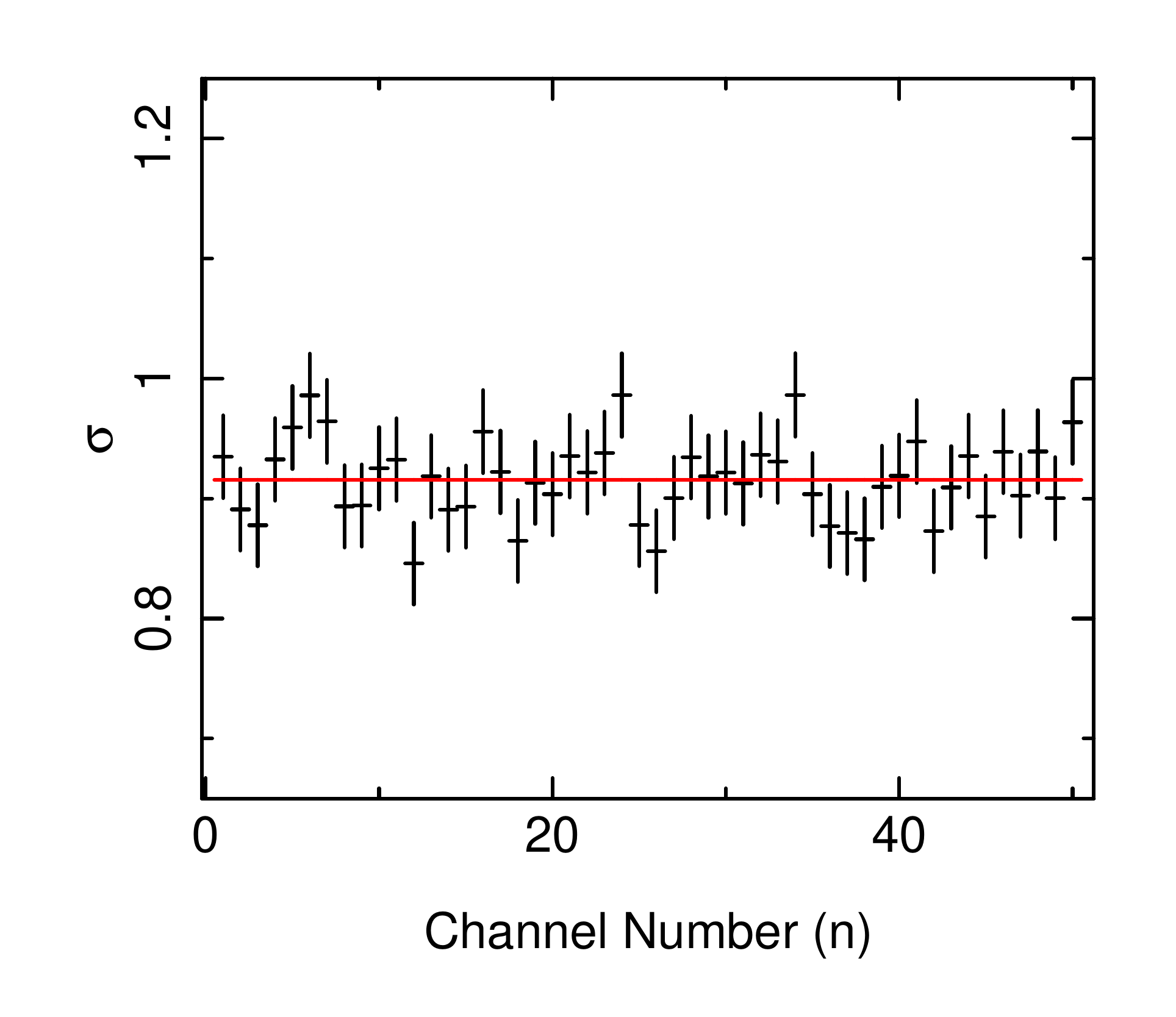} ~~~~~
    \includegraphics[width=\columnwidth,trim=0.5cm 1.0cm 1.5cm 10.0cm,clip=true]{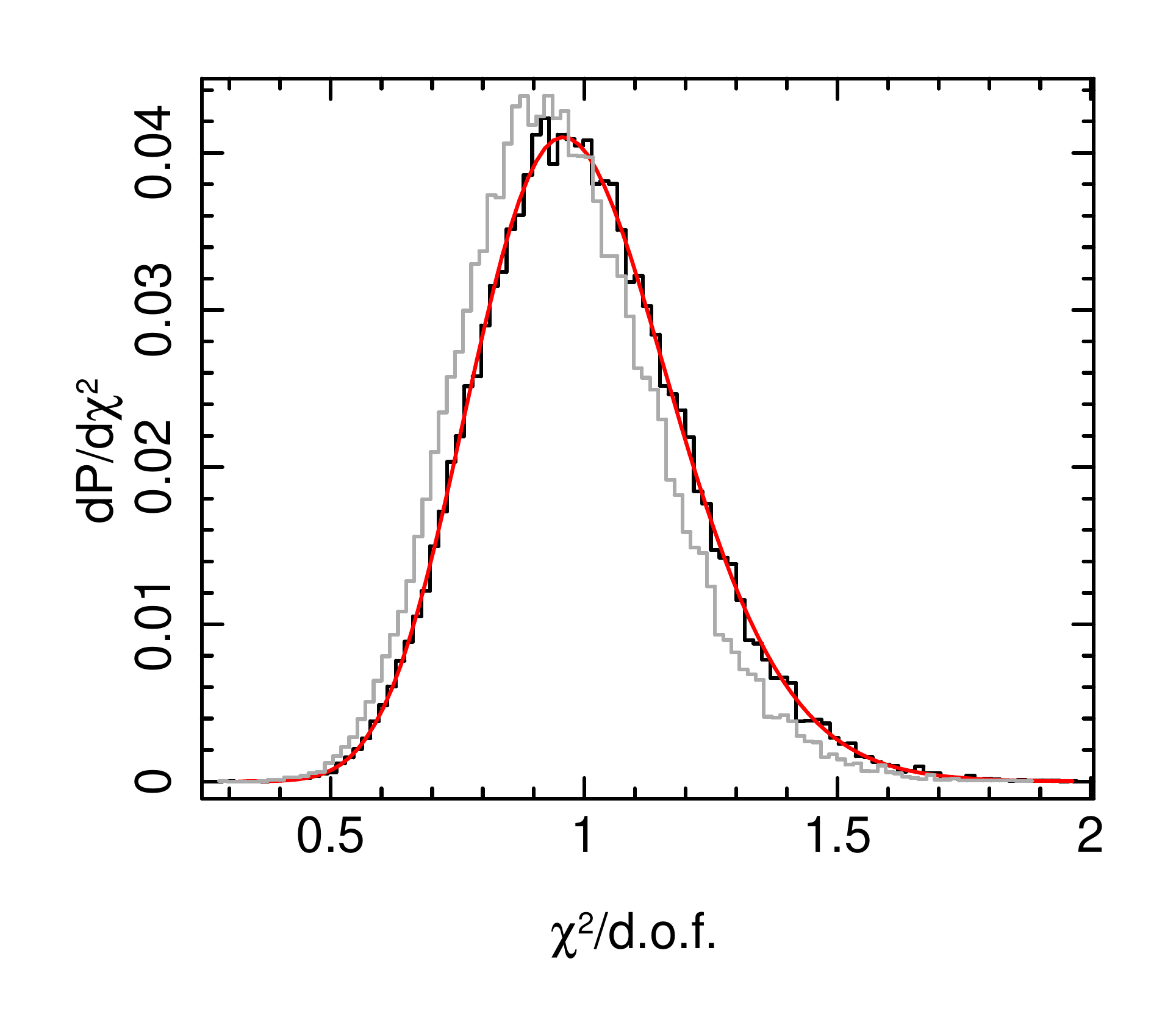}
	\vspace{-6mm}
    \caption{\textit{Left:} Simulations of an RMS spectrum with error bars calculated using equation (\ref{eqn:drms}) (black), along with the best-fitting model (red lines). \textit{Right:} Probability distributions of the $\chi^2$ value calculated for each of 50,000 such simulations. Black and grey stepped lines respectively correspond to errors calculated using equation (\ref{eqn:drms}) and the Bendat \& Piersol equivalent. The input parameters are the same as those used for Fig. \ref{fig:eg}.}
    \label{fig:egrms}
\end{figure*}

\begin{figure}
	\vspace{-6mm}
	\includegraphics[width=\columnwidth,trim=0.5cm 1.0cm 1.5cm 0.0cm,clip=true]{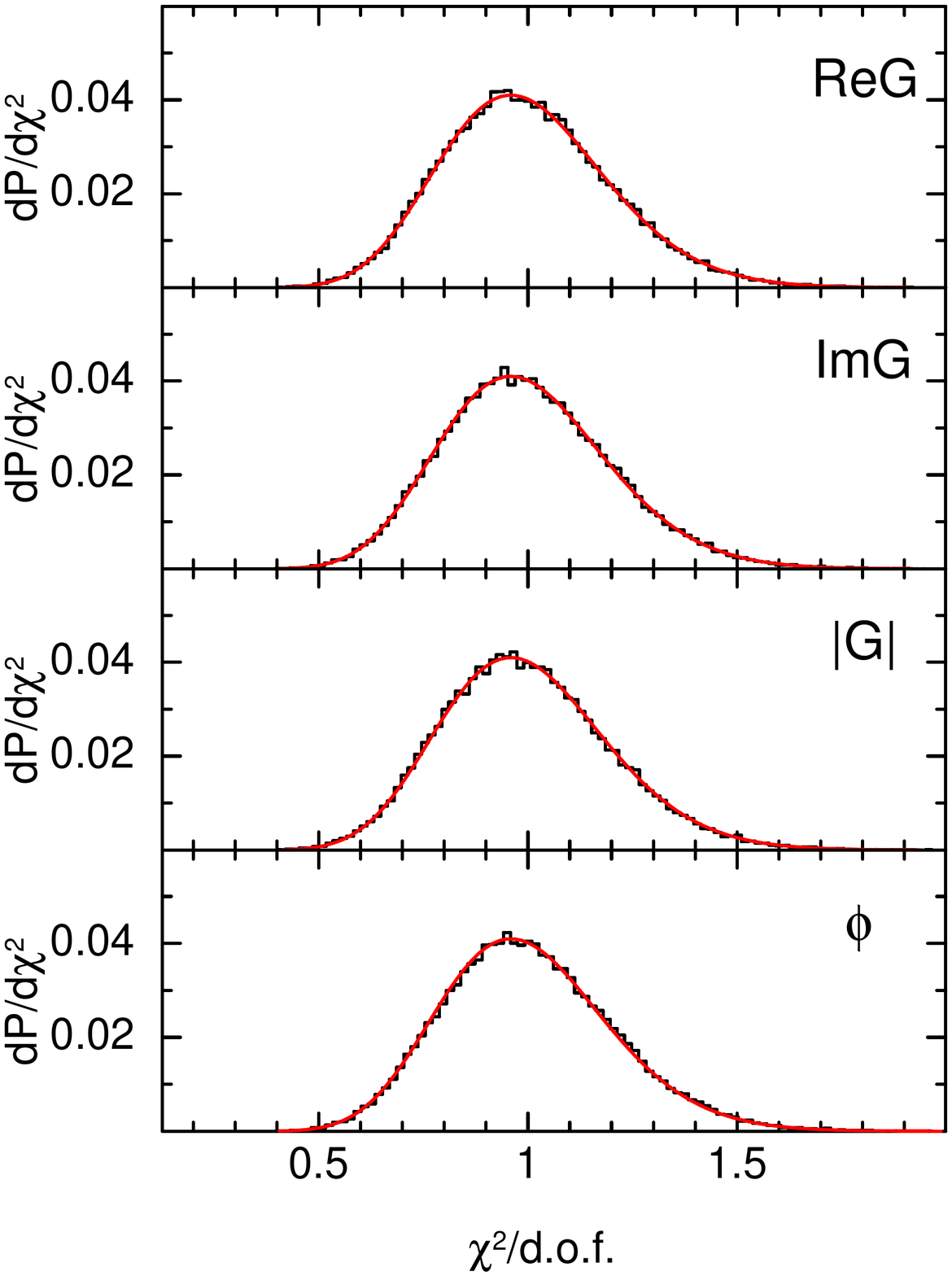}
	\vspace{-6mm}
    \caption{The same as Fig. \ref{fig:eg} except now the reference band in the simulation is the sum of all the subject bands. The error formulae derived here therefore also apply in the case whereby the reference band light curve is a sum of some or all subject band light curves.}
    \label{fig:broad}
\end{figure}

We can now use this Monte Carlo simulation to test the error formulae presented in Section \ref{sec:formulae}. I average ensemble averaged quantities over $N=500$ realizations and calculate the cross-spectrum for $N_e=50$ subject band channels, considering only one frequency range (I will therefore leave frequency dependencies as implicit for the remainder of this section). I use arbitrarily chosen input parameters: $\gamma(n)=0.6$, $R=0.6$, $P_{\rm r,noise}=0.1$, $S(n)=0.2 + i~0.9$ and $P_{\rm s,noise}(n)=0.6$ to calculate $N_e$ simulated cross-spectral estimates, $\tilde{G}(n)$, and $N_e$ estimates of the rms, $\tilde{\sigma}(n)$. Since the chosen input underlying quantities do not depend on energy channel $n$, any scatter in the simulated cross-spectra will be due purely to statistical fluctuations. Fig. \ref{fig:eg} (left) shows the simulation output. Top to bottom panels show respectively the real, imaginary, modulus and phase of the cross-spectrum, and the error bars are calculated using the formulae presented in Section \ref{sec:crosserrors}. The red lines are best-fitting constants. It is important to note that these red lines are \textit{not} exactly equal to the input models. This is because of the systematic error introduced by the reference band uncertainty. It is possible to incorporate this uncertainty into the uncertainty of the model normalization, since we know the errors on the reference band power spectrum. However, most of the time the model normalization is not of much physical interest.

The error bars in Fig. \ref{fig:eg} (left) describe the scatter of the data well, resulting in reduced $\chi^2$ values of $\sim$unity. I further test the error formulae by running 50,000 such simulations, each time with a different random seed, and measuring the same four $\chi^2$ values for each run. Fig. \ref{fig:eg} (right) shows histograms of these 50,000 $\chi^2$ values (black stepped lines) for the real part, imaginary part, modulus and phase of the cross-spectrum (top to bottom). These histograms are described very well by the $\chi^2$ probability density function with $N_e-1=49$ degrees of freedom (red lines), indicating that the error formulae are correct. The grey stepped lines result when the \cite{Bendat2010} error formulae are used instead of the new formulae presented here. Clearly, these formulae result in smaller reduced $\chi^2$ values than those predicted by the $\chi^2$ probability density function, indicating over-fitting. Corresponding results for the rms spectrum are shown in Fig. \ref{fig:egrms}. Again we see that the new formula (black stepped lines) gives the expected answer, whereas the \cite{Bendat2010} formula results in (albeit very mild) over-fitting\footnote{Note that the \cite{Bendat2010} formulae work very well for simulations of a single frequency dependent power and cross-spectrum, as expected.}.

The simulation results presented thus far all assume that the reference band is completely separate from the subject bands. I additionally ran an alternative set of simulations with the reference band instead consisting of the sum of all the subject bands. The results of this simulation are shown in Fig. \ref{fig:broad}, demonstrating that the new error formulae presented in this paper are also valid in the case whereby the reference band light curve is created by summing some or all of the subject band light curves.

\section{Derivations}
\label{sec:deriv}

I use asymptotic limits of the Monte Carlo simulation to derive analytic estimates for the errors on the energy-dependent cross-spectrum. As discussed in the previous sub-section, the Monte Carlo simulations use 4 seed complex Gaussian random variables: $X_{jk}$, $A_{jk}$, $Y_{jkn}$ and $B_{jkn}$. Here, subscript $j$ represents the $j^{\rm th}$ realization, subscript $k$ the $k^{\rm th}$ frequency range and subscript $n$ the $n^{\rm th}$ energy channel. Crucially, new values of $X_{jk}$ and $A_{jk}$ are drawn only for every frequency range of every realization (i.e. every permutation of $j$ and $k$), whereas new values of $Y_{jkn}$ and $B_{jkn}$ are drawn for every new frequency range \textit{and} channel of every realization (i.e. every permutation of $j$, $k$ and $n$).

This subtlety introduces two definitions of expectation value: one associated with averaging over all frequency ranges and the other associated with averaging over all energy channels. Taking
\begin{equation}
\langle |X_{jk}|^2 \rangle \equiv \frac{1}{N} \sum_{j=1}^N |X_{jk}|^2
\end{equation}
as an example of a random variable whose asymptotic properties do not depend on $k$ or $n$, the former and latter definitions of expectation are 
\begin{equation}
{\rm E }_k\{ \langle |X_{jk}|^2 \rangle \} = \frac{1}{N_f}\sum_{k=1}^{N_f} \langle |X_{jk}|^2 \rangle,
\end{equation}
and
\begin{equation}
{\rm E }_n \{ \langle |X_{jk}|^2 \rangle \} =\frac{1}{N_e}\sum_{n=1}^{N_e} \langle |X_{jk}|^2 \rangle.
\end{equation}
Consequently, there are two different definitions of variance: that associated with changing $k$ and that associated with changing $n$
\begin{eqnarray}
{\rm Var }_k\{ \langle |X_{jk}|^2 \rangle \} &=& {\rm E }_k \left\{ \left( \langle |X_{jk}|^2 \rangle \right )^2 \right\} - \left( {\rm E }_k \{ \langle |X_{jk}|^2 \rangle \} \right)^2 \\
{\rm Var }_n\{ \langle |X_{jk}|^2 \rangle \} &=& {\rm E }_n \left\{ \left( \langle |X_{jk}|^2 \rangle \right)^2 \right\} - \left( {\rm E }_n \{ \langle |X_{jk}|^2 \rangle \} \right)^2.
\end{eqnarray}
Hereafter, I shall refer to the former and latter cases respectively as \textit{the k variance} and \textit{the n variance}. The $k$ variance of $\langle |X_{jk}|^2 \rangle$ is
\begin{equation}
{\rm Var }_k\{ \langle |X_{jk}|^2 \rangle \} = \frac{1}{N},
\label{eqn:Xjj}
\end{equation}
which comes from the formula for the standard error on the mean of a Gaussian distributed random variable. The $n$ variance, on the other hand, is
\begin{equation}
{\rm Var }_n\{ \langle |X_{jk}|^2 \rangle \} = 0,
\label{eqn:Xj}
\end{equation}
because the same value of $X_{jk}$ is used for all energy channels, for a given realization and frequency range.

In the \cite{Bendat2010} limit {(i.e. the calculation of a single power spectrum or cross-spectrum)} the $k$ variance is the relevant quantity, whereas in the energy-dependent limit, the $n$ variance is instead what matters. The following asymptotic limits will also be important for the derivations in the coming sub-sections
\begin{eqnarray}
{\rm Var}_n \{ |Y_{jkn}|^2 \} &=& \frac{1}{N} \label{eqn:Yj} \\
{\rm Var}_n \{ {\rm Re } [X_{jk} Y_{jkn}^*] \} &=& \frac{1}{2N} \label{eqn:XjYj} \\
{\rm Var}_n \{ {\rm Re } [B_{jkn} Y_{jkn}^*] \} &=& \frac{1}{2N} \label{eqn:BjYj},
\end{eqnarray}
and random variables with the same subscripts can trivially be interchanged. Since the expectation value of any deterministic quantity is simply itself, the above relations enable me to derive all error formulae. For instance,
\begin{equation}
    {\rm Var}_n\{ |S(k,n)|^2 \langle |Y_{jkn}|^2 \rangle \} = \frac{|S(k,n)|^4}{N}.
\end{equation}

\subsection{Real and imaginary parts}
\label{sec:dGderiv}

From equations (\ref{eqn:rawG}), (\ref{eqn:Rj}) and (\ref{eqn:Sjn}), we see that the real part of the estimated cross-spectrum is equal to
\begin{equation}
    \begin{split}
{\rm Re} [\tilde{G}(k,n)] = {\rm Re}[G(k,n)] \langle |X_{jk}|^2 \rangle \\
+ \sqrt{1-\gamma^2(k,n)} {\rm Re}[S(k,n) R^*(k) \langle Y_{jkn} X_{jk}^* \rangle ] \\
+ \sqrt{P_{\rm s,noise}(k,n) P_{\rm r,noise}(k)} {\rm Re} [\langle B_{jkn} A_{jk}^* \rangle] \\
+ \gamma(k,n) \sqrt{P_{\rm r,noise}(k)} {\rm Re} [S(k,n) \langle X_{jk} A_{jk}^* \rangle ] \\
+ \sqrt{1-\gamma^2(k,n)} \sqrt{P_{\rm r,noise}(k)} {\rm Re}[ S(k,n) \langle Y_{jkn} A_{jk}^* \rangle] \\
+ \sqrt{P_{\rm s,noise}(k,n)} {\rm Re} [R^*(k) \langle B_{jkn} X_{jk}^* \rangle].
    \end{split}
    \label{eqn:ReG}
\end{equation}
Using equations (\ref{eqn:Xj})-(\ref{eqn:BjYj}), the $n$ variance of the six terms on the right hand side (RHS) of equation (\ref{eqn:ReG}) is
\begin{eqnarray}
{\rm Var}_n \{ {\rm term~1} \} &=& 0 \label{eqn:term1} \\
{\rm Var}_n \{ {\rm term~2} \} &=& \frac{[1-\gamma^2(k,n)] |S(k,n)|^2 |R(k)|^2 }{2N} \label{eqn:term2} \\
{\rm Var}_n \{ {\rm term~3} \} &=& \frac{P_{\rm s,noise}(k,n) P_{\rm r,noise}(k)}{2N} \label{eqn:term3} \\
{\rm Var}_n \{ {\rm term~4} \} &=& 0 \label{eqn:term4} \\
{\rm Var}_n \{ {\rm term~5} \} &=& \frac{[1-\gamma^2(k,n)] |S(k,n)|^2 P_{\rm r,noise}(k)}{2N} \label{eqn:term5} \\
{\rm Var}_n \{ {\rm term~6} \} &=& \frac{P_{\rm s,noise}(k,n) |R(k)|^2 }{2N}. \label{eqn:term6}
\end{eqnarray}
In the limit of $N \gtrsim 40$ explored here, all six terms are Gaussian distributed and therefore the real part of the cross-spectrum is also Gaussian distributed. Therefore, the variance of the real part of the cross-spectrum is the sum of the variances of the six terms, giving
\begin{equation}
    {\rm Var}_n \{{\rm Re}[\tilde{G}(n)]\} = \frac{\tilde{P}_r}{2N} \left[ [1-\gamma^2(n)] \tilde{P}_s(n) + \gamma^2(n) P_{\rm s,noise}(n)  \right],
\end{equation}
where the $k$ dependence has been left as implicit for brevity (and will continue to be hereafter, except for the seed random variables). The expression for the error on ${\rm Re}[\tilde{G}(n)]$ quoted in equation (\ref{eqn:dG}) results from taking the square root of the above equation and re-arranging.

The imaginary part of the cross-spectrum can also be expanded into six terms, simply by taking equation (\ref{eqn:ReG}) and substituting ${\rm Re}$ for ${\rm Im}$. The $n$ variances of these six terms are exactly the same as for the real part (equations \ref{eqn:term1}-\ref{eqn:term6}), and therefore $d{\rm Im}[\tilde{G}(n)]=d{\rm Re}[\tilde{G}(n)]$.

We can use the same logic to derive the \cite{Bendat2010} formulae for real and imaginary parts of the cross-spectrum. In this case, there is only one value of $n$, and we wish to know the $k$ variance, which is the same as the $n$ variance for terms 2, 3, 5 and 6. The $k$ variance of term 4 is ${\rm Var}_k\{ {\rm term~4}\} = \gamma^2(n) |S(n)|^2 P_{\rm r,noise}/(2N)$ for both the real and imaginary parts of the cross-spectrum. For term 1, it is $({\rm Re}[G(n)])^2/N$ for the real part and $({\rm Im}[G(n)])^2/N$ for the imaginary part. Summing the variance of the six terms, re-arranging and taking the square root then leads to the \cite{Bendat2010} error formulae (equations \ref{eqn:BPdG}).

\subsection{Modulus and phase}
\label{sec:dphideriv}

Since the errors on the real and imaginary parts of the energy-dependent cross-spectrum are equal to one another, we can trivially write $d|\tilde{G}(n)|=d{\rm Im}[\tilde{G}(n)]=d{\rm Re}[\tilde{G}(n)]$. This is simple to appreciate if we picture error contours on the complex plane. The distribution is a Gaussian in the real and imaginary axes, and therefore the error contours will simply be concentric circles if the errors on the real and imaginary axes are the same. Therefore, the width of the distribution is the same in any direction on the complex plane.

Using geometrical arguments, \cite{Bendat2010} show that the variance of the phase angle can be written as
\begin{equation}
\begin{split}
{\rm Var}\{\tilde{\phi}\} &=& \frac{1}{|G|^4} \Big\{
({\rm Re}[G])^2 {\rm Var}\{{\rm Im} [\tilde{G}] \} \\
& & -2 {\rm Re}[G] {\rm Im}[G] {\rm Cov}\{{\rm Re} [\tilde{G}],{\rm Im} [\tilde{G}] \} \\
& &+ ({\rm Im}[G])^2 {\rm Var}\{{\rm Re} [\tilde{G}]\}
\Big\},
\end{split}
\label{eqn:dphi1}
\end{equation}
and this is valid both for the $k$ and $n$ variance. Here, the covariance term is
\begin{equation}
\begin{split}
    {\rm Cov}\{{\rm Re}[\tilde{G}(n)],{\rm Im}[\tilde{G}(n)]\}
    = {\rm E}\{ {\rm Re}[\tilde{G}(n)] {\rm Im}[\tilde{G}(n)] \} \\
    - {\rm E}\{ {\rm Re}[\tilde{G}(n)] \} {\rm E}\{ {\rm Im}[\tilde{G}(n)] \}.
    \end{split}
    \label{eqn:covdef}
\end{equation}
We can calculate ${\rm Re} [\tilde{G}(n)]{\rm Im} [\tilde{G}(n)]$ by taking equation (\ref{eqn:ReG}) and multiplying it by the equivalent expression for the imaginary part of the cross-spectrum to get
\begin{equation}
    {\rm Re} [\tilde{G}(n)]{\rm Im} [\tilde{G}(n)] = {\rm Re}[G(n)] {\rm Im}[G(n)] (\langle |X_{jk}|^2 \rangle)^2 + \rm{35~more~terms}.
    \label{eqn:cov}
\end{equation}
Substituting all 36 terms on the RHS of equation (\ref{eqn:cov}) into equation (\ref{eqn:covdef}) at first appears a rather daunting task. We can however take a shortcut  by first considering the $k$ covariance, which we already know to be 
${\rm Cov}_k\{{\rm Re} [\tilde{G}],{\rm Im} [\tilde{G}] \}={\rm Re}[G]{\rm Im}[G]/N$ \citep{Bendat2010}. We can see that this is equal to the $k$ covariance of the first term on the RHS of equation (\ref{eqn:cov}). We can therefore conclude that the other 35 terms contribute zero $k$ covariance. Therefore, all we need to do to calculate the covariance term in the energy-dependent limit is evaluate the $n$ variance of the first term on the RHS of equation (\ref{eqn:cov}), which is zero. Therefore, ${\rm Cov}_n\{{\rm Re} [\tilde{G}],{\rm Im} [\tilde{G}] \}=0$.

With the covariance term equal to zero, and the variances of the real and imaginary parts equal to one another, equation (\ref{eqn:dphi1}) dramatically simplifies to
\begin{equation}
    {\rm Var}_n \{\tilde{\phi}(n)\} = \frac{{\rm Var}_n \{{\rm Re} [\tilde{G}(n)]\}}{|G(n)|^2},
\end{equation}
and it is simple to take the square root and re-arrange to arrive at equation (\ref{eqn:dphi}).






\subsection{RMS spectrum}
\label{sec:dpsderiv}

Substituting equation (\ref{eqn:Sjn}) into equation (\ref{eqn:Ps}) and multiplying out of the brackets gives
\begin{equation}
    \tilde{P}_s(n) = \gamma^2(n) |S(n)|^2 \langle |X_{jk}|^2\rangle + {\rm 8~more~terms}.
    \label{eqn:9terms}
\end{equation}
For the first term, the $n$ variance is zero, whereas for the other 8 terms, the $n$ variance is equal to the $k$ variance. Therefore, using a similar trick to the previous sub-section, we can simply start from the \cite{Bendat2010} formula for the variance on a single power spectrum, ${\rm Var}_k \{ \tilde{P}_s \}=\tilde{P}_s^2/N$, and subtract the $k$ variance of the first term in equation (\ref{eqn:9terms}), which is
\begin{equation}
    {\rm Var}_k \{ \gamma^2(n) |S(n)|^2 \langle |X_j|^2\rangle \} = \frac{\gamma^4(n)|S(n)|^4}{N}.
\end{equation}
This gives
\begin{equation}
    {\rm Var }_n \{ \tilde{P}_s(n) \} = \frac{ \left[ |S(n)|^2 + P_{\rm s,noise} \right]^2 - \gamma^4(n)|S(n)|^4}{N},
\end{equation}
which is straight forward to manipulate into equation (\ref{eqn:dPs}).

\section{Observational example}
\label{sec:data}

\begin{figure}
	\vspace{-6mm}
	\includegraphics[width=\columnwidth,trim=0.5cm 1.0cm 1.5cm 10.0cm,clip=true]{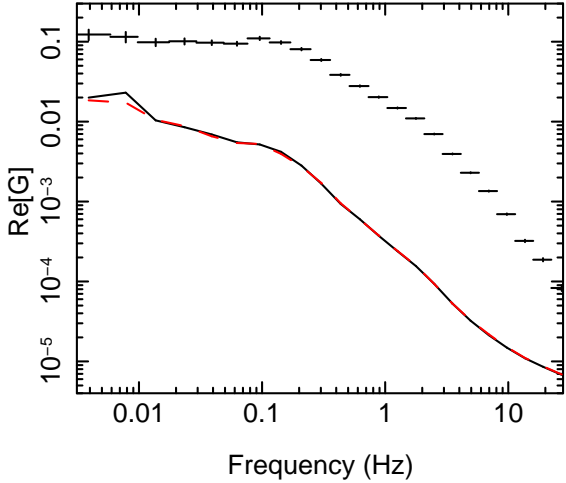}
	\vspace{-6mm}
    \caption{Example of the real part of the cross-spectrum for \textit{RXTE} PCA data from Cygnus X-1 (black points). The reference band is $0.14~{\rm keV} \leq E \leq 4.34~{\rm keV}$ and the subject band is $4.34~{\rm keV} \leq E \leq 4.64~{\rm keV}$. The error bars are calculated from the Bendat \& Piersol formulae (equation \ref{eqn:BPdG}). The black solid and red dashed lines show the errors calculated directly from the standard error on the mean and with the Bendat \& Piersol formula respectively.}
    \label{fig:cygx1}
\end{figure}

I demonstrate the use of the new formulae on an \textit{RXTE} observation of Cygnus X-1 in the hard state (obsid 10238-01-06-00). This observation has been used for a number of spectral-timing studies (e.g. \citealt{Gilfanov2000,Mastroserio2018,Mahmoud2018}). Most significantly for the purposes of this paper, \cite{Mastroserio2019} fit the X-ray reverberation model \textsc{reltrans} \citep{Ingram2019} to the real and imaginary parts of the energy-dependent cross-spectrum for 10 frequency ranges (plus the time-average spectrum) in order to yield a black hole mass measurement, but the best fitting model had a reduced $\chi^2$ of $0.76$. It turns out that this over-fitting occurred because the error bars were calculated in the single cross-spectrum limit relevant to the \cite{Bendat2010} formulae, when the error formulae derived in this paper would instead have been appropriate.

I follow the data reduction procedure described in \cite{Mastroserio2018} to obtain $N_e=64$ light curves each with time bin sizes of $dt=1/64$ s (the best resolution possible for this particular binned mode). I ensemble average the power and cross spectra over 46 segments, each of $256$ s duration, resulting in a useful exposure of $11.776$ ks. I apply geometric re-binning to the power and cross spectra, {such that the $k^{\rm th}$ frequency bin is averaged over $N_f(k) \le c_0^k$ Fourier frequencies. Here, $N_f$ is an integer and $c_0 \ge 1$ is a constant. Therefore, for large $k$ the width of the frequency bins increases logarithmicaly with frequency}. An accurate estimate of the
noise {component} in the reference band and all subject bands is required. Since the noise is significantly affected by dead time in this data set (due to the high count rate), and there is no frequency range below the Nyquist frequency that is dominated by noise (in which case the noise level could have been determined empirically), it is necessary to use a dead time model. I use the \cite{Zhang1995} dead time model of the Proportional Counter Array (PCA). I ignore very large events, which only become important for frequencies much higher than the Nyquist frequency of this data set (which is $32$ Hz). Specifically, I use equation 4 from \cite{Nowak1999} with very large event rate $\mathcal{R}_{\rm vle}=0$, PCA dead time $\tau_d=10~\mu$s, time bin size $t_b=dt=1/64$ s and the number of proportional counter units (PCUs) is 5. This is a fairly general formula, and so in principle can be applied for other {X-ray} telescopes as long as the dead time is known, or the Nyquist frequency is high enough to fit the model to a noise dominated frequency range (see e.g. \citealt{Bult2017} for the case of \textit{NuSTAR}).

In order to calculate the bias term, $\tilde{b}^2$ (equation \ref{eqn:b2}), I set up an iteration loop. The loop is initialized by calculating $\tilde{b}^2$ in the $\gamma^2=1$ limit, and this is then used to calculate $\tilde{\gamma}^2$ (equation \ref{eqn:g2}). This value of $\tilde{\gamma}^2$ is then used to calculate a new value of $\tilde{b}^2$, which in turn is used to calculate another new value of $\tilde{\gamma}^2$ and so on until convergence is achieved. This typically takes only a few steps, and the results turn out to be very similar to simply calculating $\tilde{b}^2$ in the $\gamma^2=1$ limit because the coherence is very close to unity for most frequencies of most bands. For some very low count rate energy bands (with $E \gtrsim 50$ keV well out of the calibrated range of the PCA), $\tilde{b}^2$ is larger than $|\tilde{G}|^2$ due to the large uncertainties involved in making the estimates. Whenever this is the case, I simply set $\tilde{b}^2=0$. This does not affect any of the bands used for the analysis.

\begin{figure}
	\vspace{-6mm}
	\includegraphics[width=\columnwidth,trim=1.0cm 1.0cm 1.5cm 10.0cm,clip=true]{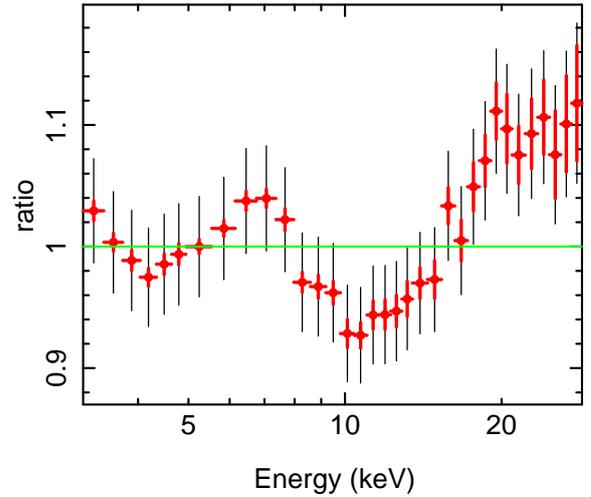}
	\vspace{-6mm}
    \caption{Example of the real part of the energy-dependent cross-spectrum for \textit{RXTE} PCA data from Cygnus X-1 (black points) in the frequency range $0.041-0.092$ Hz, plotted as a ratio to an absorbed power-law model folded around the PCA response matrix. The black points have error bars calculated from the Bendat \& Piersol formulae, and the red points use the new formula derived in this paper. The error bars calculated from the old formula are clearly too large, whereas the new error bars properly reflect the scatter seen in the data.}
    \label{fig:cygx1E}
\end{figure}

The black points in Fig. \ref{fig:cygx1} depict the real part of the cross-spectrum between a subject band consisting of channel 10 ($4.34$ - $4.64$ keV) and a reference band consisting of channels 1-9 ($0.14$ - $4.34$ keV). I use fractional rms normalization, re-bin using a geometric binning constant of $c_0=1.4$ and the error bars are calculated for the single cross-spectrum limit (i.e. equation \ref{eqn:BPdG}), since this is a single cross-spectrum plotted against frequency. The black line depicts the error on ${\rm Re}[\tilde{G}]$ estimated directly from the standard error around the mean. For each Fourier frequency, this is calculated as the standard deviation of the 46 estimates of the real part of the cross-spectrum (i.e. one for each light curve segment) divided by $\sqrt{46}$. The errors of all the Fourier frequencies in each broad frequency bin are then averaged in quadrature. The red dashed line is the error instead calculated from equation (\ref{eqn:BPdG}). We see that this gives the same answer, as expected. This indicates that the use of independent Gaussian random variables in the derivation of the error formulae does not render them inaccurate for use on data from accreting compact objects, even though we know that the real data exhibit non-linear variability properties not captured by the \cite{Timmer1995} algorithm \citep{Uttley2005,Maccarone2011}.

Fig. \ref{fig:cygx1E} shows the real part of the energy-dependent cross-spectrum in the frequency range $0.041$ - $0.092$ Hz. The reference band is now the full PCA band (i.e. the sum of all 64 subject bands), the normalization is now \textit{absolute} rms and the re-binning constant is now $c_0=1.9$. Note that, because each subject band is included in the reference band, a noise contribution, $\mathcal{N}(n)$, has been subtracted from each subject band (see Section \ref{sec:refband}). This noise contribution is always very small compared to the signal because there are many more counts in the reference band than in any given subject band. The data are plotted as the ratio to an absorbed power law model (\textsc{tbabs} with $N_h=6\times 10^{21} {\rm cm}^{-2}$ and the \citealt{Wilms2000} abundances, $\Gamma=1.852$) that has been folded around the PCA response matrix. The black and red points respectively use the error formula for the single cross-spectrum (equation \ref{eqn:BPdG}) and energy-dependent (equation \ref{eqn:dG}) limit. We see that the error bars are clearly too large in the former case, whereas they are commensurate with the spread of the data around broad trends in the latter case. It is worth noting that this is an extreme example chosen for illustration. For higher frequencies of the real part and all frequencies of the imaginary part, the discrepancy between the two formulae is much less striking. Since the error bars in \cite{Mastroserio2019} were calculated from the standard error on the mean (i.e. the \citealt{Bendat2010} limit), this explains the mild over-fitting exhibited by their analysis: the over-fitting will have been fairly severe for some spectra in the simultaneous fit, with little to no over-fitting for other spectra. {Fortunately}, changing the error formulae is unlikely to have a large systematic effect on the best fitting parameters, because the low frequency cross-spectrum (which is most affected by over-fitting) and the time-averaged spectrum predicted by the reverberation model are very similar to one another, and the time-averaged spectrum is very well constrained by the data. In particular, the black hole mass is most sensitive to the signal at higher frequencies. The uncertainties on the model parameters may however have been over-estimated.

\begin{figure}
	\vspace{-6mm}
	\includegraphics[width=\columnwidth,trim=1.0cm 1.0cm 1.5cm 6.0cm,clip=true]{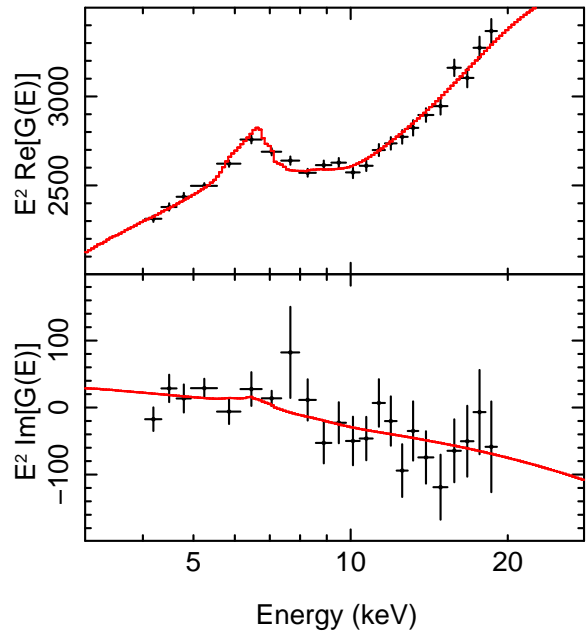}
	\vspace{-6mm}
    \caption{Example fit of the X-ray reverberation model \textsc{reltrans} to the $0.041-0.092$ Hz energy-dependent cross-spectrum. Real and imaginary parts are in the top and bottom panels respectively. The data are unfolded around the best fitting model (\textsc{xspec} command \texttt{iplot eeuf}). The reduced $\chi^2$ is 42.9/36, which corresponds to a null-hypothesis probability of $20\%$.}
    \label{fig:reltrans}
\end{figure}

As a final test of the formulae, I simultaneously fit \textsc{reltrans} to the real and imaginary parts of the $4-20$ keV energy-dependent cross-spectrum in the frequency range $0.041-0.092$ Hz only, using \textsc{xspec} version 12.9 \citep{Arnaud1996}. This is not indented as a means to constrain system parameters, but rather as an exercise to test whether a reasonable $\chi^2$ can be achieved using the new error formulae. The result is shown in Fig. \ref{fig:reltrans}, with the real and imaginary parts of the energy-dependent cross-spectrum plotted in the top and bottom panels respectively. The data are unfolded around the best fitting model (\textsc{xspec} command \texttt{iplot eeuf}) and the model is plotted with a higher resolution than the data for clarity (\textsc{xspec} command \texttt{iplot eemo}). To obtain this fit, I started from the Model 3 parameters of \cite{Mastroserio2019}. I then varied the `continuum' parameters $\alpha$, $\gamma$, $\phi_A$ and $\phi_B$ whilst keeping the other parameters fixed. This is necessary because I use a different reference band to \cite{Mastroserio2019}, and I use the cross-spectrum instead of the complex covariance, meaning that the continuum parameters have slightly different definitions. I then additionally thaw the parameters $h$, $r_{\rm in}$, $\Gamma$, $\log\xi$, $A_{\rm Fe}$ and $1/\mathcal{B}$. The new best fitting values of these parameters are consistent with the \cite{Mastroserio2019} parameters within $1~\sigma$ confidence. The resulting reduced $\chi^2$ is $42.9/36$, which corresponds to a null-hypothesis probability of $20\%$, indicating a formally acceptable fit.

The Fortran code used for this section can be downloaded from \url{https://bitbucket.org/adingram/cross_spec_code/downloads/}.

\section{Discussion}
\label{sec:discussion}

I have presented error formulae for the energy-dependent cross-spectrum and rms spectrum, and shown that these formulae are still valid when a single broad reference band is used. Here, I discuss {the} results {of this work}.

\subsection{The optimal reference band}

I have shown that a broad reference band can be used to calculate the energy-dependent cross-spectrum, as long as a noise contribution is subtracted from the real part of the cross-spectrum for subject bands that are part of the reference band. {For photon counting noise,} the noise contribution tends to be very small compared with the signal, since the count rate in the broad reference band will always be much greater than that in an individual subject band. It is common in the literature to avoid trivial correlations between the subject and reference bands by subtracting the current subject band from the broad reference band, meaning that each subject band is crossed with a slightly different reference band. This only introduces a small bias, again because the reference band count rate is much larger than the subject band count rate, but the method suggested here is far more elegant and introduces no bias at all as long as the noise level is well known.

With freedom to define the reference band however we see fit, we can ask the question: what is the optimal reference band? In the most general case, we can define the reference band light curve as a weighted sum of all the subject band light curves. We could then define the weightings in order to minimize e.g. $\sum_n^{N_e} d|\tilde{G}(n)|/|\tilde{G}(n)|$ (see equation \ref{eqn:dG}). However, this would become a rather involved process, and to a good approximation signal to noise is more or less optimized by maximizing the reference band count rate. Therefore, for most practical applications, it is optimal for the reference band to be the sum of all the subject bands. 
{A possible exception is if the phase normalization of the model that will be fit to the data is calculated self-consistently,}
in which case it is optimal to restrict the reference band only to well-calibrated energy channels (see \citealt{Mastroserio2018} and \citealt{Ingram2019} for a detailed discussion).

\subsection{When to use which formula?}

For the case of a model being fit to the frequency dependence of a single power and/or cross-spectrum (e.g. \citealt{Rapisarda2017,Rapisarda2017a}), the \cite{Bendat2010} formulae should be used to calculate error bars. For the case of a model being fit to the energy dependence of the power and/or cross-spectrum (including the time / phase lags) in a single frequency range (e.g. \citealt{Cackett2014,Chainakun2016}), {the new} formulae {presented in this work} should be used to calculate error bars. {Similarly}, if a model is being fit to the energy-dependent power and/or cross-spectrum for a number of frequency ranges, \textit{and the model normalization is a free parameter for different frequency ranges}, {the new} formulae should be used. 
{This is because, in this case, the fit to the energy-dependent cross-spectrum for each frequency range is completely independent of the fit to the energy-dependent cross-spectrum in every other frequency range. The formula for a single frequency range therefore applies.} This is most certainly the case for any \textit{frequency resolved spectroscopy} studies, where spectral models are fit to the energy-dependent rms or covariance spectrum for different frequency ranges (e.g. \citealt{Revnivtsev1999,Gilfanov2000,Axelsson2013,Peille2015,Axelsson2018}). This is also case for the \textsc{reltrans} model, since {in this model the amplitude of the energy-dependent cross-spectrum in the frequency range centered on frequency $\nu$ is set by the normalization parameter $\alpha(\nu)$, which is} free to take any value for each frequency range in the fit\footnote{Although note that in a plot of best fitting $\alpha$ against frequency, the error bars will look too small because the uncertainty in $\alpha$ contributed by the reference band is effectively treated by the new formulae as a systematic error rather than a statistical error.}. {The} new formulae should also be used for the X-ray polarimetry-timing method of \cite{Ingram2017a}, meaning that scope for detecting X-ray polarization QPOs is a little more positive than the analysis presented in that paper suggests.

However, if a model predicts the frequency \textit{and} energy dependence of the cross-spectrum and/or power spectra, then the \cite{Bendat2010} formulae should be used. This is because the normalization of the energy-dependent model is now not free to take any value that minimizes $\chi^2$ for a given frequency range. This is the case for the propagating mass accretion rate model \textsc{propfluc} (e.g. \citealt{Ingram2012,Ingram2013,Rapisarda2016}), which has so far only been applied to a single cross-spectrum and the power spectra of two bands, but could in principle be extended to multiple energy bands. This would also be the case if the \textsc{reltrans} model were changed such that a frequency dependent model for $\alpha(\nu)$ were implemented (e.g. a zero-centered Lorentzian).

\subsection{Comparison with other formulae}

I have extensively compared the formulae derived here with the \cite{Bendat2010} formulae, and discussed the limits in which each set of equations should be used. It is important to also discuss other formulae in the literature that cover slightly different limits.

\subsubsection{RMS Spectrum}

The error on the rms spectrum has previously been derived in the $\gamma=1$ limit, first empirically from Monte Carlo simulations by \cite{Vaughan2003} and later analytically by \cite{Wilkinson2011a}. Taking equation (\ref{eqn:drms}) {from this paper} and setting $\gamma=1$ yields an equation almost identical to the $\gamma=1$ limit formula quoted in \cite{Uttley2014} (their equation 14), except their equation has a $2$ in the denominator instead of a $4$. This turns out to result from a very subtle mistake made when converting from the time domain excess variance considered in \cite{Vaughan2003} (equation 11 therein) and \cite{Wilkinson2011a} to the frequency domain rms considered in \cite{Uttley2014}. The variance in the earlier two papers is calculated in the time domain for $N_t$ time bins (this is called $N$ in \citealt{Vaughan2003} and \citealt{Wilkinson2011a}, but I have re-named it in {an} attempt to avoid confusion). This corresponds, via Parseval's theorem, to the integral of the power spectrum over all $N_t/2$ positive Fourier frequencies.
In the notation of \cite{Uttley2014}, the power spectra are averaged over $K$ frequency bins and $M$ light curve segments, such that the total number of realizations being averaged over is $N=KM$. The analysis of \cite{Vaughan2003} essentially uses $K=N_t/2$ and $M=1$ (integrating over all frequency ranges but doing no ensemble averaging over light curve segments). Therefore, the formula for the error on the rms spectrum quoted in \cite{Uttley2014} (equation 14 therein) must be divided by $\sqrt{2}$ in order to be correct in the $\gamma=1$ limit.

\subsubsection{Covariance spectrum}

The error on the covariance spectrum in the $\gamma=1$ limit is also quoted in \cite{Uttley2014} (equation 15 therein). This formula was derived by \cite{Wilkinson2009} using equations from \cite{Bartlett1955}. We can compare this with formula {derived in this work} by taking equation (\ref{eqn:dG}) and setting $\gamma=1$ to get
\begin{equation}
    d|\tilde{C}(n)| = \sqrt{ \frac{ \sigma_r^2 \sigma_{\rm noise}^2(n) + \sigma_{\rm r,noise}^2 \sigma_{\rm noise}^2(n) }{ 2 N \sigma_r^2 } },
\end{equation}
where $\sigma_r^2 \equiv \Delta\nu [\tilde{P}_r - P_{\rm r,noise}]$ and $\sigma_{\rm r,noise}^2 \equiv \Delta\nu P_{\rm r,noise}$. This formula is similar to equation 15 in \cite{Uttley2014}, except their equation has an extra term in the numerator: $\sigma^2(n) \sigma_{\rm r,noise}^2$ in {the} notation {of this paper}. This extra term arises because the formulae {in \cite{Bartlett1955}} \textit{do} account for the signal from the reference band being the same for each subject band, but \textit{don't} account for the noise in the reference band being the same for every subject band. However, there is only one reference band light curve used to make the energy-dependent cross-spectrum, with only one signal contribution and one noise contribution. Equation 15 from \cite{Uttley2014} {can be reproduced}
by using ${\rm Var}_j\{ {\rm term~4 } \} = \gamma^2(n)|S(n)|^2 P_{\rm r,noise}/(2N)$ in place of ${\rm Var}_j\{ {\rm term~4 } \} =0$ (see equation \ref{eqn:term4}) in {the} derivation for the error on the real and imaginary parts of the cross-spectrum {presented in Section \ref{sec:dGderiv}}, and finally setting $\gamma=1$.

\subsection{Impact on the literature to date}

{It seems possible}
that the use of incorrect error formulae in the literature has led to many erroneous results. This is not the case though - the conclusions of previous studies are unlikely to be significantly changed by adoption of the new error formulae (see e.g. Section \ref{sec:data}), {since the new formulae are in the most part subtle corrections to the old formulae.}
That said, it is clearly optimal to use the new formulae where appropriate in future, since they are formally correct and no harder to implement than the old formulae.

There are many studies of lag versus energy spectra in the literature that use the \cite{Bendat2010} error formula (e.g. \citealt{Kara2013,Kara2016,Cackett2014,Chainakun2016}). Although this formula gives error bars that are slightly too large, Fig. \ref{fig:eg} (bottom panel) shows that the resulting over-fitting is very subtle indeed, and therefore should not be a cause for concern. In contrast, any frequency resolved spectroscopy studies that employed the \cite{Bendat2010} error formulae will have suffered from severe over-fitting (see the third panel down of Fig. \ref{fig:eg}). However, most if not all frequency resolved spectroscopy studies use either the \cite{Vaughan2003} formula for the error on the rms in the $\gamma=1$ limit or the \cite{Wilkinson2009} formula for the error on the covariance in the $\gamma=1$ limit. As discussed in the previous sub-section, there is a small mistake in the version of the \cite{Vaughan2003} formula quoted in \cite{Uttley2014}, but this is only a factor of $\sqrt{2}$ and it is not clear whether it has propagated into the literature at all. The previous sub-section also discusses an extra term in the \cite{Wilkinson2009} formula that technically should not be included, but in practice is very small if a broad reference band is used. Therefore, the error formulae applied in previous studies (e.g. \citealt{Wilkinson2009,Axelsson2014}) can all be considered approximately correct, although in future I recommend the use of the new formulae presented here.

\section{Conclusions}
\label{sec:conclusions}

I have derived new error formulae for the energy-dependent cross-spectrum and rms spectrum that are valid for {any value of} intrinsic coherence. I present errors for the energy-dependent cross-spectrum in terms of real and imaginary parts (equation \ref{eqn:dG}) and in terms of modulus (equation \ref{eqn:dG}) and phase (equation \ref{eqn:dphi}). The error on the rms spectrum is equation (\ref{eqn:drms}).

I have shown that it is optimal to use a broad reference band, constructed by summing the flux from many subject bands. In this case, trivial correlations arise between the reference band and the subject bands that are included in the reference band. It is common in the literature to avoid these trivial correlations by subtracting the current subject band count rate from the reference band, leading to a slightly different reference band being used for each subject band. Here I show that this is not necessary. Instead, the same broad reference band should be used for each subject band. If the current subject band is included in the reference band, then the trivial correlations can be accounted for by subtracting a noise contribution from the real part of the cross-spectrum.

Finally, I have discussed the general form of the bias term that arises when calculating the modulus of the cross-spectrum (equation \ref{eqn:b2}). In order to accurately calculate the intrinsic coherence, it is necessary to set up an iteration loop, since the intrinsic coherence and bias depend on one another. I find for an example of hard state Cygnus X-1 data that this iteration loop converges quickly and the measured intrinsic coherence is similar to that measured using the $\gamma=1$ limit of the bias term (which has been employed in the literature in the past). This is mainly because the intrinsic coherence is very close to unity for the observation considered.

A Fortran code that employs the new error formulae, a broad reference band and uses an iteration loop to properly calculate the intrinsic coherence can be downloaded from \url{https://bitbucket.org/adingram/cross_spec_code/downloads/}. This code was used for the Cygnus X-1 example discussed in Section \ref{sec:data}. Comprehensive use instructions can be found in the download repository.

\section*{Acknowledgements}

I thank Michiel van der Klis, Edward Nathan, Guglielmo Mastroserio and Phil Uttley for valuable discussions. I acknowledge support from the Royal Society, {and useful comments from the anonymous referee that improved the paper.}




\bibliographystyle{mnras}
\bibliography{biblio} 








\bsp	
\label{lastpage}
\end{document}